\documentclass[twocolumn,showpacs,preprintnumbers,amsmath,amssymb]{revtex4}

\usepackage{graphicx}

\begin{document}

\title{Atypical thermodynamic behavior of Ce compounds in the vicinity of zero temperature Critical Points}

\author{Julian G. Sereni}
\address{Div. Bajas Temperaturas, Centro At\'omico Bariloche (CNEA), 8400 S.C. Bariloche, Argentina}

\date{\today}

\begin{abstract}

{A systematic analysis of thermodynamic properties performed on
Ce-base exemplary compounds allows to identify different types of
behaviors as the system approaches the quantum critical region.
They are recognized in the respective magnetic phase boundaries
($T_{N,C}(x)$) as a change from the classical negative curvature
to a linear composition ($x$) dependence, the occurrence of a
critical point or the evanescence at finite temperature under
pressure at finite temperature. In the first case, an anomalous
reduction of the entropy $S_m$ respect to the $S_m=R\ln2$ value
(expected for the usual Ce-magnetic doublet ground state) is
observed around $x_{cr}$, and analyzed taking profit of detailed
studies performed on $CeIn_{3-x}Sn_x$ alloys. As expected from
Maxwell relations, the volume variation $V_0(x)$ at $T\to 0$ also
shows a non-monotonous behavior around $x_{cr}$.

Different regimes in the entropy variation of the ordered phase
($S_{MO}$) are recognized. Only in the former case $S_{MO}\to 0$
continuously as $T_{N,C} \to 0$, whereas in the second
$S_{MO}(x,B)$ remains constant till a first order transition
occurs. In the third case, the degrees of freedom of the MO phase
are progressively transferred to the heavy fermion component as
indicated by the decreasing $\Delta C_m(T_N)$ jump which vanishes
at finite temperature.}

\end{abstract}

\pacs{jsereni at cab.cnea.gov.ar} \maketitle

\section{Introduction}

Magnetic phase transitions can be experimentally driven by
external control parameters such as chemical composition, pressure
or magnetic field.  Their application allows to trace the
evolution of fluctuations related to a second order transition as
its associated thermal energy decreases. In the limit of zero
temperature, thermal fluctuations become intrinsically frozen and
a phase transition may only have a quantum character. In fact, a
quantum critical point (QCP) is defined \cite{TVojta} as the $T=0$
limit for a second order transition driven by one of the mentioned
non-thermal control parameters. Despite of its unattainable
nature, a $T=0$ QCP presents a sort of 'halo' of related quantum
fluctuations whose physical effects are observed at finite
temperature \cite{HvL,Steglich}. The phenomenology arising from
those low lying energy excitations is known as that of a
'non-Fermi-liquid' (NFL) \cite{HvL,Steglich,Stewart}, in contrast
to the canonical Fermi-liquid observed in non-magnetic regimes.
One of the most relevant feature of NFL systems is the increasing
density of low energy excitations reflected as a divergency of
thermal parameters like specific heat divided temperature
($C_m/T$) when $T\to 0$. Other physical parameters, like magnetic
susceptibility, thermal expansion and electrical resistivity also
show unusual behaviors \cite{Stewart}.

In this work, we focus on the thermodynamical implications the
peculiar behavior of $C_m/T$ and the related entropy ($S_m$) at
low temperature. The phase boundaries taken into account are
driven by the current control parameters: chemical potential,
pressure and magnetic field. In the former case, the chemical
potential is tuned by the variation of the composition of
Ce-ligands. The related experimental evidences are analyzed in
Section II and discussed in Section III. These results are
compared in Section IV with the phenomenology observed in other
types of phase diagrams including pressure driven systems.

\section{Characteristic properties of Ce-ligand alloyed systems.}

Three exemplary Ce binary compounds are taken as referents for the
present analysis after their deeply investigated low temperature
properties. In all of them the ordering temperatures ($T_{N,C}$)
were driven within an extended range by changing the composition
of Ce-ligands ($x$). This procedure allows to preserve the lattice
of Ce magnetic atoms without changes in the local symmetry and
minimizing any modification of the magnetic interactions. Their
respective magnetic behaviors are antiferromagnetic (AF):
$CeIn_{3-x}Sn_x$ \cite{Pedraza} and $CePd_2Ge_{2-x}Si_x$
\cite{Octavio}, and ferromagnetic (FM): $CePd_{1-x}Rh_x$
\cite{CePdRh}.

\begin{figure}
\begin{center}
\includegraphics[angle=0, width=0.5 \textwidth] {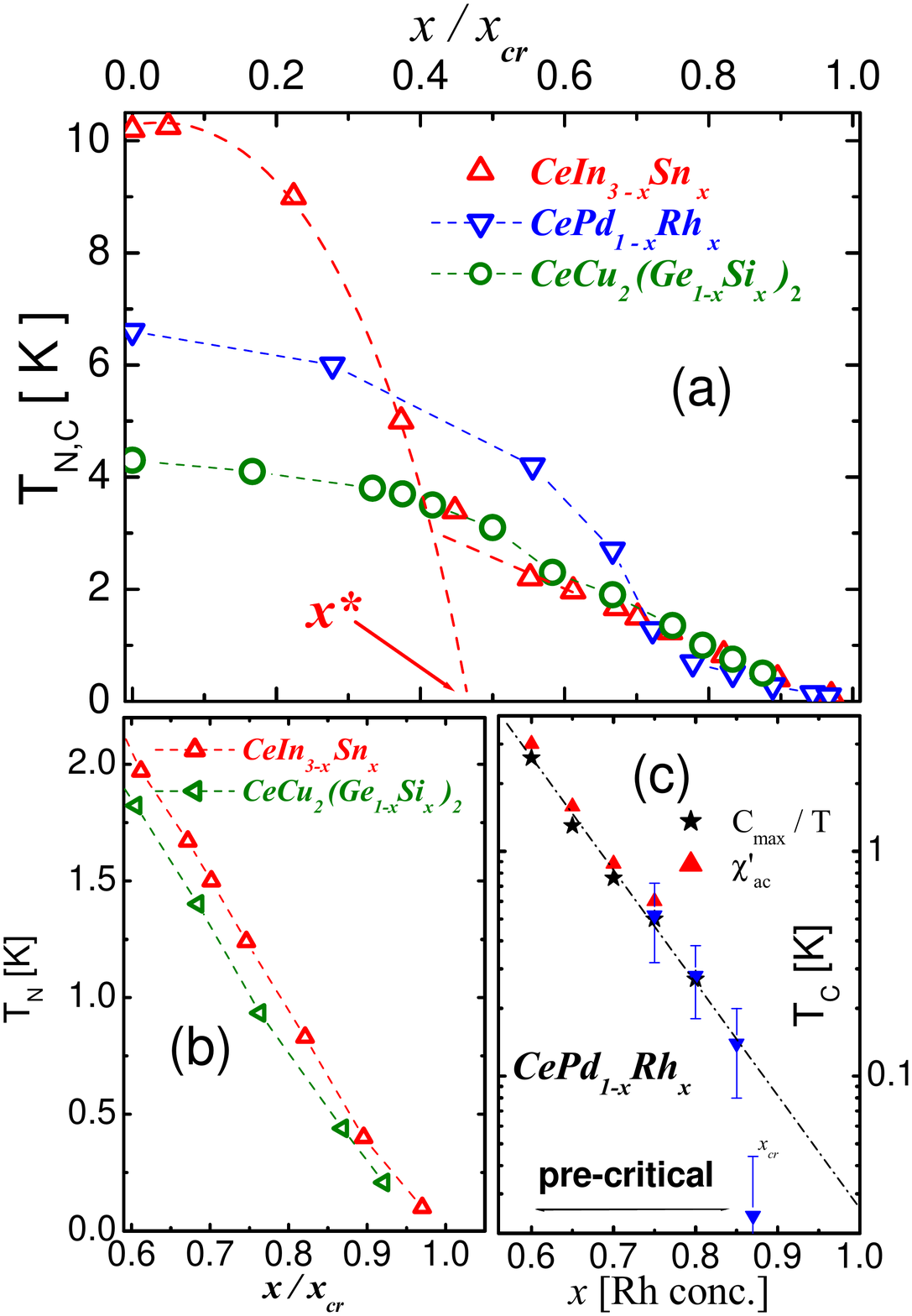}
\end{center}
\caption{(Color online) (a) Magnetic phase diagrams of three
exemplary compounds covering an extended range of temperature. (b)
Detail of the pre-critical region of two AF compounds with linear
$T_N(x)$ dependence. (c) Asymptotic phase boundary of the FM
compound. Notice the logarithmic $T_C$ axis. x* indicates the
extrapolation for $CeIn_{3-x}Sn_x$ from the classical region }
\label{F1}
\end{figure}

A typical magnetic phase boundary related to a second order
transition shows a negative curvature as a function of applied
control parameters. Such a phase boundary extrapolates to a
critical value $x^*$ as the transition temperature $T_{N,C}\to 0$,
see e.g. the case of $CeIn_{3-x}Sn_x$ in Fig.~\ref{F1}a. In the
mentioned exemplary compounds, however, a change of curvature
around 2K occurs \cite{anivHvL} as shown in Fig.~\ref{F1}a. At
that concentration the system enters into a pre-critical region
between $x^*<x<x_{cr}$. That change of regime is also observed in
other Ce-ligand concentration dependent systems but, to our
knowledge, not in pressure or magnetic field driven systems.

The observed change of regime can be explained by taking into
account the competition between the decreasing energy of the
thermal fluctuations (which extrapolates to $x^*$) and the
temperature independent energy of the quantum fluctuations.
Despite of the fact that quantum fluctuations are related to $T\to
0$ phenomena, these experimental evidences indicates that below
$\approx 2$\,K the latter mechanism takes over and dominates the
scenario driving $T_{N} \propto \mid x-x_{cr} \mid$ as can be
observed in Fig.~\ref{F1}b for two AF exemplary compounds. In the
case of the FM one (see Fig.~\ref{F1}c) $T_{C}(x)$ decreases
asymptotically till it collapses to zero at a first order
transition. Strictly, the alternative of a linear $T_C(x)$
dependence in the very last concentrations before $x_{cr}$ can be
considered as discussed in Ref. \cite{CePdRh}.

\begin{figure}
\begin{center}
\includegraphics[angle=0, width=0.5 \textwidth] {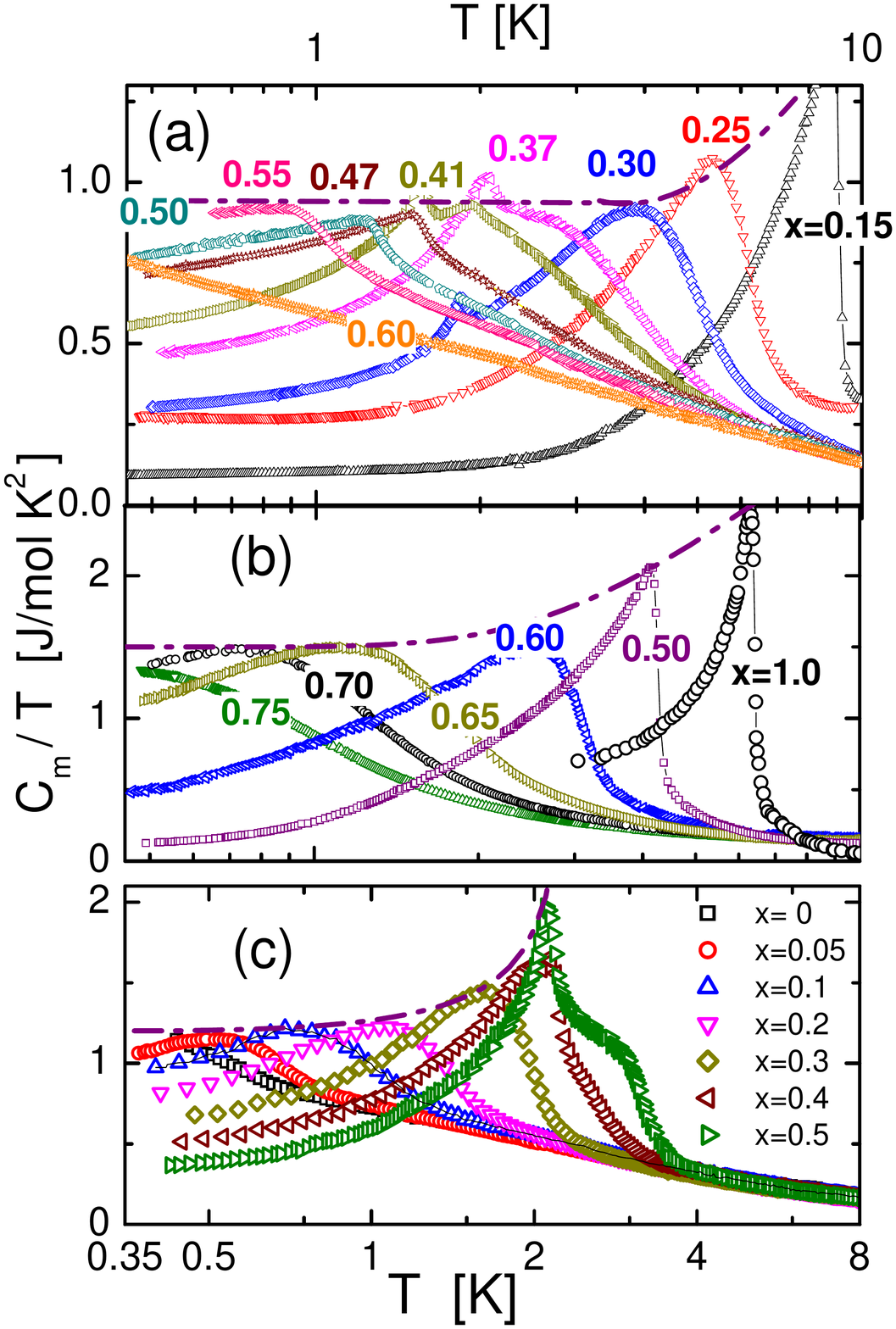}
\end{center}
\caption{(Color online) Low temperature specific heat divided
temperature of three exemplary compounds approaching their
respective pre-critical regime: (a) AF- $CeIn_{3-x}Sn_x$, (b)
FM-$CePd_{1-x}Rh_x$ and (c) AF-$CePd_2(Ge_{1-x}Si_x)_2$. Dash-dot
curves remark the nearly constant value of the $C_m/T$ maxima
within the pre-critical region.} \label{F2}
\end{figure}

\subsection{Specific heat}

Within the pre-critical region, the specific heat of the exemplary
compounds show a common temperature dependence since the
respective maxima $C_{max}(T_{N,C})/T$ tends to a constant value
\cite{anivHvL} as the critical concentration is approached, see
Fig.~\ref{F2}. Hereafter, $C_m$ indicates the magnetic
contribution to the specific heat after phonon subtraction
extracted from the respective La isotypic compounds. Such a
behavior for $x > x^*$ clearly differs from the observed within
the canonical regime ($x < x^*$) where $C_{max}(T_{N,C})/T$
decreases with $T_{N,C}(x)$. This behavior can be analyzed within
the scope of the Ginzburg-Landau theory for a second order
transition, where $C_m/T=a^2/2b$ at $T_{N,C}$, being $a$ and $b$
the coefficients of the free energy expansion $G(\psi,T)=G_0(T)+
a(T)\psi^2+ b(T)\psi^4$. This indicates that, within the
pre-critical region, the $G(\psi,T)$ dependence on the $a^2/b$
ratio is locked and consequently the entropy of the ordered phase
decreases linearly according to the law of corresponding states.

\begin{figure}
\begin{center}
\includegraphics[angle=0, width=0.52 \textwidth] {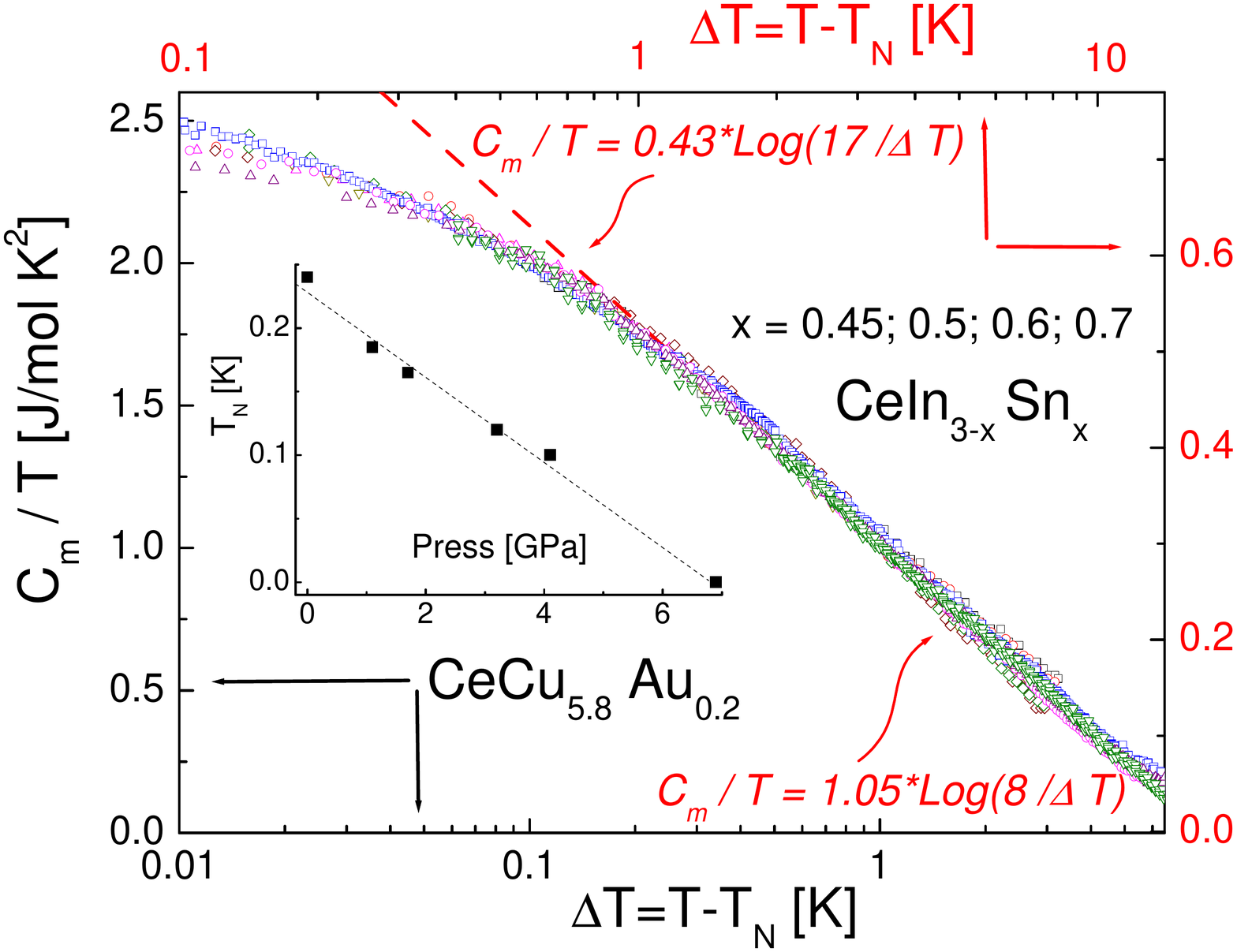}
\end{center}
\caption{Color online) Comparison between two AF systems with
linear $T(x)$ dependence versus their respective $\Delta T =
T-T_N$ temperatures in a $C_m/T$ vs. $log T$ representation. Data
for $CeCu_{5.8}Au_{0.2}$ under pressure was extracted from
\cite{HvLPietrus}. Inset: linear $T_N$ dependence on pressure.}
\label{F3}
\end{figure}

Further peculiarities are observed in the tail of $C_m(T)/T$ above
$T_{N,C}$ which shows a divergent $-log(T/T_0)$ dependence
characteristic of NFL systems \cite{scaling}. This behavior is
illustrated in Fig.~\ref{F3} for two AF compounds. To remark the
universality of this logarithmic dependence, we compare in that
figure the $C_m(T)/T$ results for different concentrations of
$CeIn_{3-x}Sn_x$ within the pre-critical concentration range and
different applied pressures ($p$) on $CeCu_{5.8}Au_{0.2}$ which
lies close to the critical point \cite{HvLPietrus}. Since in both
cases $T_N$ decreases linearly with respective control parameters
$x$ and $p$ their $C_m(T)/T$ tails can be scaled by a simple shift
of the temperature like $\Delta T = T - T_N$. Notice the low
temperature flattening of $C_m/T$ which skips the $T\to 0$
divergence according to thermodynamic laws.

\subsection{Low temperature properties of $CeIn_{3-x}Sn_x$}

\begin{figure}
\begin{center}
\includegraphics[angle=0, width=0.5 \textwidth] {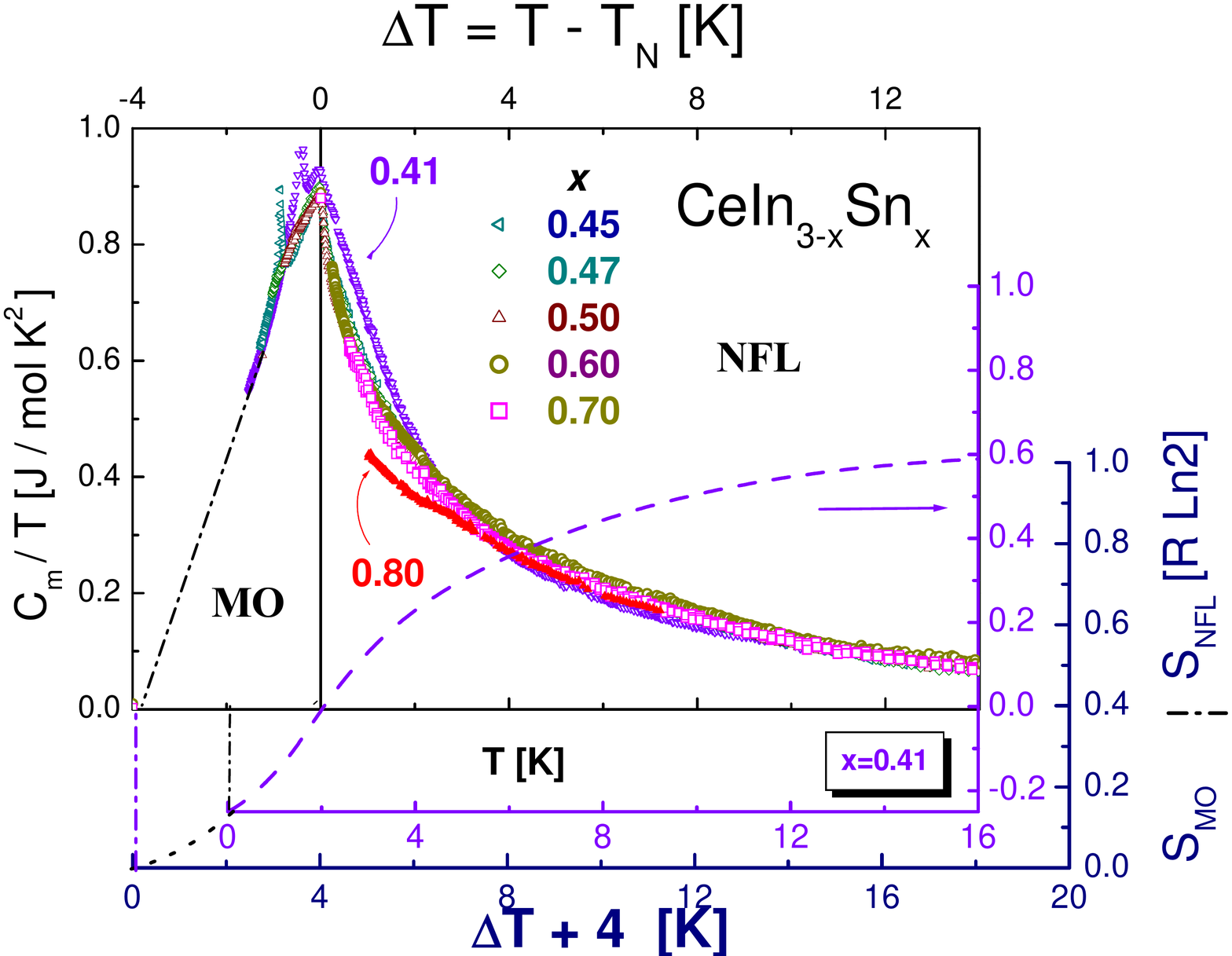}
\end{center}
\caption{(Color online) Overlap of $C_m/T (0.45\leq x \leq 0.70)$
curves plotted vs.\,a shifted temperature $\Delta T=T-T_N$ (upper
and left axes) after Ref.\cite{anivHvL}. The dashed curve
represents the unique entropy gain vs. T for $0.45\leq x \leq
0.70$ samples, discriminated between ordered $S_{M0}$ and
paramagnetic $S_{NFL}$ contributions (low $T$ and inner $S$ axes).
Full $RLn2$ entropy is computed from $\Delta T=-4$K (lowest
`$\Delta T+4$' and outer `$S_{MO}+S_{NFL}$' right axes), see the
text.} \label{F4}
\end{figure}

In this section we will analyze the low temperature magnetic
contribution to the specific heat and entropy ($S_m(T)$) of
$CeIn_{3-x}Sn_x$, which was investigated in detail around its
critical concentration \cite{Pedraza}. Similarly to the analysis
performed in Fig.~\ref{F3}, we present in Fig.~\ref{F4} the
$C_m(T)/T$ dependence extended to seven concentrations ($0.41\leq
x \leq 0.80$) versus the previously defined normalized temperature
$\Delta T = T - T_N$, now including the magnetically ordered phase
into the negative range of $\Delta T$, see the upper x-axis.
There, one can see the already mentioned scaling of the
$C_m(T>T_N)/T$ tails of the alloys belonging to the pre-critical
region ($0.45\leq x \leq 0.70$). In order to remark the validity
of this scaling, we also include in that figure the results
obtained from the $x=0.41$ and $0.80$ alloys placed beyond limits
of the pre-critical region. As it can be seen, their respective
$C_m/T$ dependencies clearly deviate from the scaled ones.

In Fig.~\ref{F4}, the vertical line at $\Delta T = 0$ line splits
the $C_m/T$ contribution into two parts, i) one corresponding to
the ordered phase (MO), hereafter label as $C_{MO}/T$ and ii) the
tail at $\Delta T > 0$, hereafter identified as $C_{NFL}/T$
because of its NFL behavior. Notably, also the $C_{MO}/T$
contributions for the samples within the pre-critical region
overlap each other in this representation. Samples $x=0.41$ and
0.45 show a weak peak slightly below $T_N$, related to a first
order transition appearing around $x=x^*$, but playing no role in
the present study. The relevant feature is that the $C_{MO}/T$
overlap allows an extrapolation of $C_{MO}/T \to 0$ to $\Delta T
\approx - 4$\,K which is independent of concentration. We remark
that a $\Delta T < 0$ value does not correspond to a negative
temperature but simply to a common extrapolation to a zero value
of the $C_m /T$ contribution.

The key parameter to describe this peculiar behavior of the
specific heat is its associated entropy, evaluated as
$S_m=\mathbf{\int} C_m/TdT$. According to the definition proposed
for $\Delta T$, one may split the total entropy (c.f. measured
$S_m(T)$) as $S_m=S_{MO}+S_{NFL}$, being $S_{MO}$ the contribution
of the MO phase for $\Delta T\leq 0$ and $S_{NFL}$ the one from
the NFL tail for $\Delta T \geq 0$. For the following analysis we
take as reference the entropy variation of sample $x = 0.41$
because it contains largest $S_{MO}$ contribution among the
samples included in Fig.~\ref{F4}. As it can be appreciated in the
figure, the $S_{MO}(x=0.41)$ contribution slightly exceeds
$0.2RLn2$ whereas $S_{NFL}$ reaches $\approx 0.6RLn2$ (see inner
right axis). Noteworthy, the full $RLn2$ value is only reached
once the extrapolation to the $C_m/T = 0$ value at $\Delta T
\approx -4$\,K is included, as depicted using the lowest `$\Delta
T+4$' and outer `$S_{MO}+S_{NFL}$' right axes in Fig.~\ref{F4}.
Since $S_{NFL} \approx 0.6RLn2$ does not change with
concentration, but $S_{MO}\to 0$ as $x \to x_{cr}$ one concludes
that about $40\%$ of the $RLn2$ entropy is missed as $T_N \to 0$.

It is evident from Figs.~\ref{F3} and ~\ref{F4} that the decrease
of $S_{MO}(x\to x_{cr})$ as $\Delta T \to 0$ is not transferred to
the NFL phase at $\Delta T > 0$ because $S_{NFL}$ is independent
of concentration in this concentration range. The relevant
conclusion of this analysis is that the degrees of freedom become
exhausted at $x=x_{cr}$ whereas those belonging to the NFL phase
remain unchanged on the $60\%$ value of $RLn2$.

The loss of entropy showed by this Ce compound is not an exception
because, in the cases where this type of analysis was performed,
it was found that the R$Ln2$ value for a doublet ground state (GS)
is not reached. Particularly, the compounds showing a $C_m/T
\propto log(T/T_0)$ dependence cannot not exceed $\approx 60\%$
R$Ln2$ \cite{scaling}. Since the mentioned $log (T/T_0)$
dependence corresponds to one of the possible scenarios for QCPs
predicted by theory \cite{HvL}, it means that this lack of entropy
or the consequent arising of {\it zero point entropy} ($S_0$) is
intrinsic to the NFL phenomenology approaching that point.
Simplistic explanations looking for a some extra entropy
contribution at higher temperature fail because it would imply a
discontinuous transference of entropy from the $T_N\to 0$ MO phase
to temperature above 20\,K according to Fig.~\ref{F4}. We recall
that in CeIn$_3$ crystal-field excited quartet lie high enough in
energy ($\approx 100$\,K \cite{Lawrance}) to not be involved in
the present analysis.

This decrease of entropy at a fixed temperature around the
critical concentration can neither be attributed to the Kondo
temperature increase because, in such a case, the proper T scaling
would have been $t(x)=T/T_K(x)$ instead of the observed $\Delta T
= T - T_N$. A direct experimental evidence that $T_K$ practically
does not change within that range of concentration is given by the
fixed temperature of the maximum of the electrical resistivity
($T^{\rho}_{max}$). In CeIn$_{3-x}$Sn$_x$ \cite{Pedraza}
$T^{\rho}_{max}\approx 19K \approx T_0$ within the pre-critical
region, and it starts to increase only beyond $x_{cr}$
\cite{Pedra2}.

\section{Discussion}

\subsection{Characteristic critical concentrations}

A full magnetic phase diagram covering all the possible GS of Ce
systems should include at least three characteristic critical
concentrations were clear changes of regime occur. To our
knowledge, the first attempt to encompassing such a phase diagram
was performed nearly two decades ago by comparing different
magnetic behaviors extracted from seventeen Ce systems driven by
Ce-ligands alloying \cite{systemat94}. That analysis covered all
possible Ce GS from local moment (magnetic) regime to the unstable
valence (non-magnetic) one, and allowed to recognized two relevant
concentrations: one related to the zero temperature extrapolation
of the magnetic phase boundary (previously identified as $x^*$),
and the other where the paramagnetic temperature starts to rise
powered by the increase of the Kondo screening ($x_K$). Different
types of phase diagrams were recognized depending whether $x^*<
x_K$ or $x^*\geq x_K$. A third characteristic concentration was
identified in Ce systems reaching the unstable valence regime at
$x_{UV}$. This concentration is related to the appearance of the
sixfold degenerated Fermi liquid behavior, originated in the Ce
$J=5/2$ Hund's rule angular moment. This characteristic
concentration exceeds the purpose of the present study because we
are limited to the twofold GS regime.

Since at that time the usual low temperature limit for magnetic
studies was around one degree Kelvin, no quantum fluctuation
effects were evident enough to be taken into account. Therefore,
the $0 \leq x \leq x^*$ range dominated by thermal fluctuations
and showing the typical negative curvature of $T_{N,C}$ was taken
as valid down to $T=0$. The new available experimental information
indicates that approaching $x^*$ there is a change of curvature
(presente in Fig.~\ref{F1}), occurring at finite temperature
$T_{N,C}\approx 2$\,K. At present, the $x \to x^*$ extrapolation
is currently left aside because the $T_{N,C}\to 0$ limit occurs at
higher concentration ($x_{cr} > x^*$). Nevertheless, $x^*$ keeps
its relevance because it identifies a characteristic concentration
at which the decreasing thermal energy would have driven the phase
boundary to $T=0$ in the absence of quantum (i.e. non thermal)
phenomena.

\begin{figure}
\begin{center}
\includegraphics[angle=0, width=0.5 \textwidth] {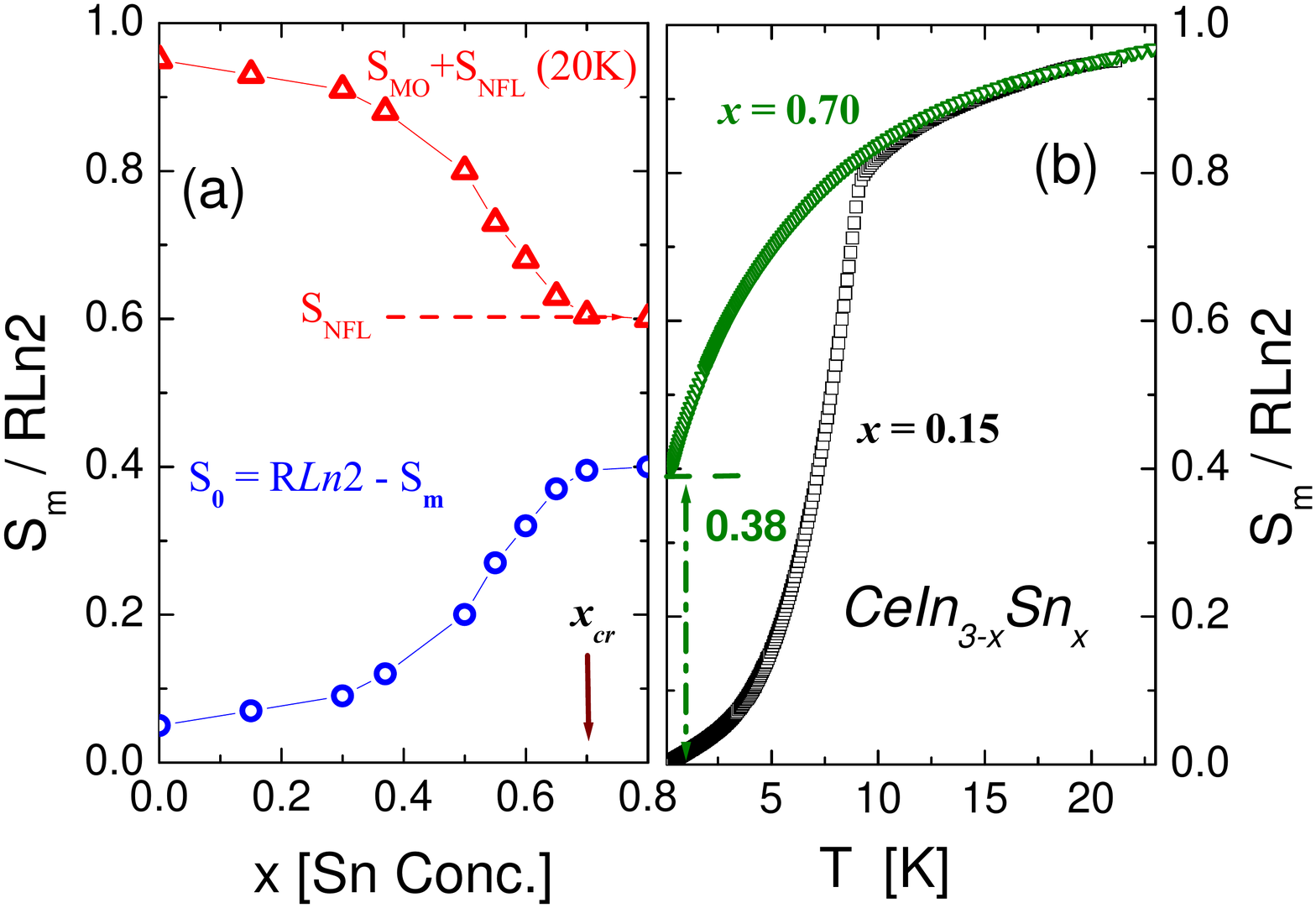}
\end{center}
\caption{(Color online) a) Concentration dependence of the entropy
$S_m(x)$ of $CeIn_{3-x}Sn_x$ measured up to 20K and the zero point
entropy $S_0$ computed as the difference respect the total
expected value R$Ln2$. The arrow indicate the critical
concentration. b) Comparison of the temperature dependence of
$S_m$ between an alloy $x=0.15$ belonging to the classical region
and one lying on top of the critical concentration $x=0.70$
showing a deficit of $\approx 40\%$.} \label{F5}
\end{figure}

Apart from the mentioned modification of the $T_{N,C}(x)$
curvature, the change of regime around $x^*$ is related to some
interesting features occurring at that concentration. Among them,
there is the formation of a new phase at $T_I \leq T(x^*)$ like in
$CeIn_{3-x}Sn_x$ \cite{Pedraza} and $CePd_2Ge_{2-x}Si_x$
\cite{Octavio}. Moreover, if we compare these phase diagrams with
those driven by applied pressure, one sees that the corresponding
$p^*$ value of that control parameter is frequently at the edge of
the appearance of superconductivity \cite{superc}. Notably, in
that case there is no change of curvature in $T_N(p)$ because the
phase boundary itself vanishes above the superconductive dome.

\subsection{Pre-critical region and Zero point Entropy}

Focusing now in the $x^* \leq x \leq x_{cr}$ region, the
outstanding effect to be discussed is the striking reduction of
the entropy respect to the $RLn2$ value. In Fig.~\ref{F5}a, we
show the variation of the total entropy with $x$ split into the
two components $S_m = S_{MO}+S_{NFL}$ corresponding to the ordered
and paramagnetic phases and computed up to 20\,K. The progressive
suppression of $S_{MO}$ as $x \to x_{cr}$ clearly contrasts with
the fixed thermal variation of $S_{NFL}$. This is a rare case
where the missed entropy can be attributed to a zero point entropy
$S_0$ because at $x=x_{cr}$ $S_{MO}=0$. To show the contrast
between this and the classical regime, we compare in
Fig.~\ref{F5}b the $S_m(T)$ variation of sample $x=0.7\approx
x_{cr}$ with a representative of the pure thermal regime
$x=0.15$. The observed difference at $T=0$ is 0.38RLn2.

\begin{figure}
\begin{center}
\includegraphics[angle=0, width=0.5 \textwidth] {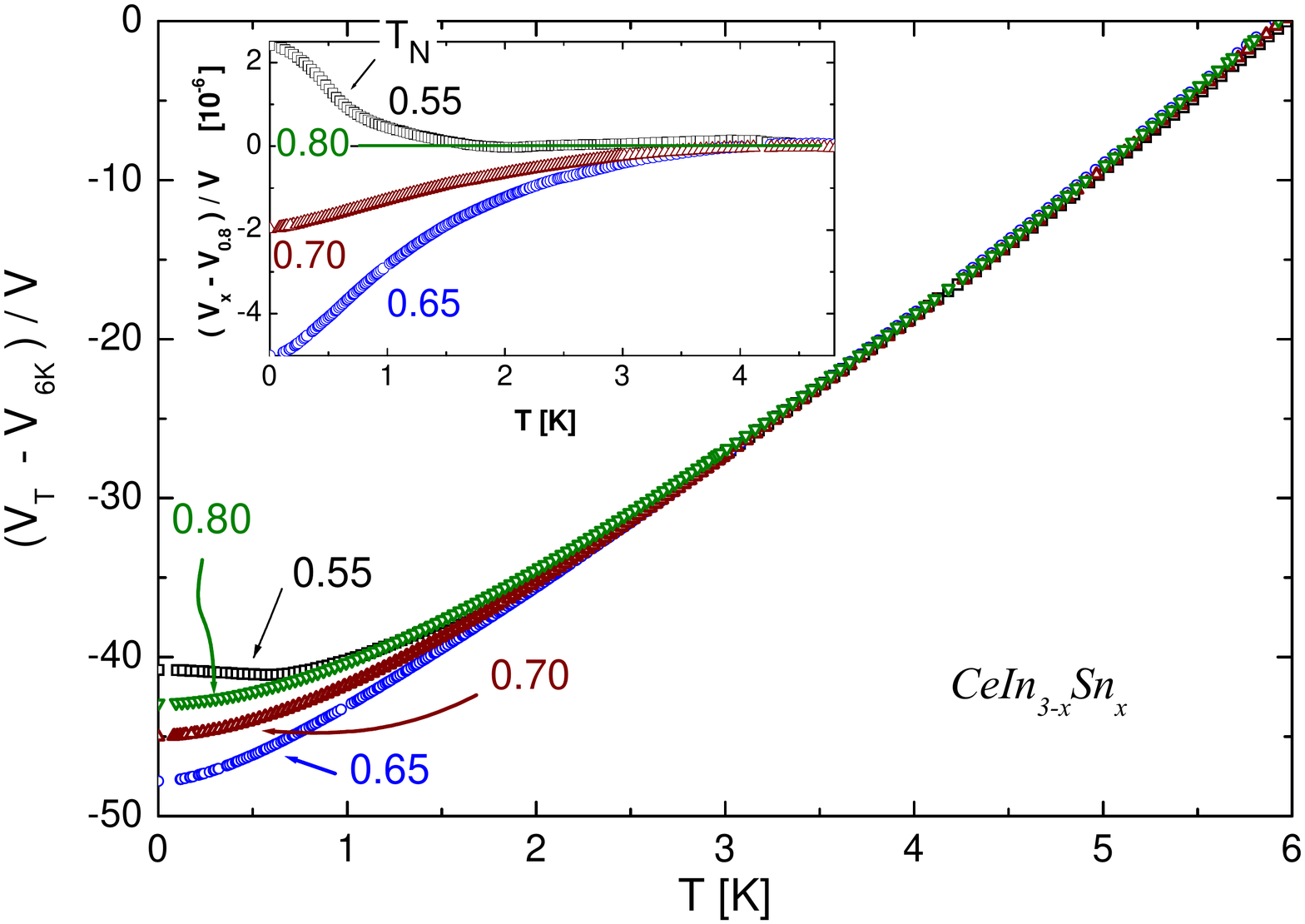}
\end{center}
\caption{(Color online) Low temperature thermal variation of the
unit cell volume normalized at $T=8$\,K, well above any quantum
fluctuation effect. Thermal expansion data from \cite{Kuechler}
Inset, detail of the $V(x,T\to 0)$ variation taking as a reference
the $x=0.80$ sample placed above the critical concentration.}
\label{F6}
\end{figure}

An alternative description for this phenomenon can be derived from
the concept of 'Rare Regions' proposed in Ref.\cite{TVojta} where
those regions can be conceived as a sort of magnetic clusters.
This scenario may apply for magnetic moments interacting FM like
in CePd$_{1-x}$Rh$_x$, however for AF systems like
$CeIn_{3-x}Sn_x$ such a cluster formation is unlikely. If one
takes into account that at $T_N\to 0$ only quantum fluctuations
may allow the system to access to two minima via quantum
tunnelling, a new degeneracy would arise as the quantum critical
regime takes over. For a quantum phase transition, those minima
would correspond to different phases which cannot be thermally
connected at $T\to 0$. In such a context, the entropy collected at
finite temperature corresponds to the progressive thermal access
to the excited Karmer's level. In such a case a $S_{NFL}=
R\ln(3/2)$ increase of entropy would be expected instead of
$R\ln2$. Quantitatively, this value corresponds to the
experimentally observed one because $\ln(3/2) = 0.6\ln 2$.

\subsection{Thermal Expansion}

In order to confirm the present analysis of the anomalous
evolution of the entropy approaching the critical point as due to
an intrinsic effect, another thermodynamic parameter sensitive to
this phenomenon has to be looked for. Such alternative is provided
by the thermal expansion $\beta(T,x)$ which is related to the
entropy through the Maxwell relation $-\partial S/\partial P =
\partial V/\partial T$.Thus an anomalous $S_0(x\to x_{cr})$ dependence should have a
replica in $V_0(x\to x_{cr})$ as $T\to 0$. In this case, the
effective pressure is originated in the {\it chemical pressure}
produced by alloying.

\begin{figure}
\begin{center}
\includegraphics[angle=0, width=0.5 \textwidth] {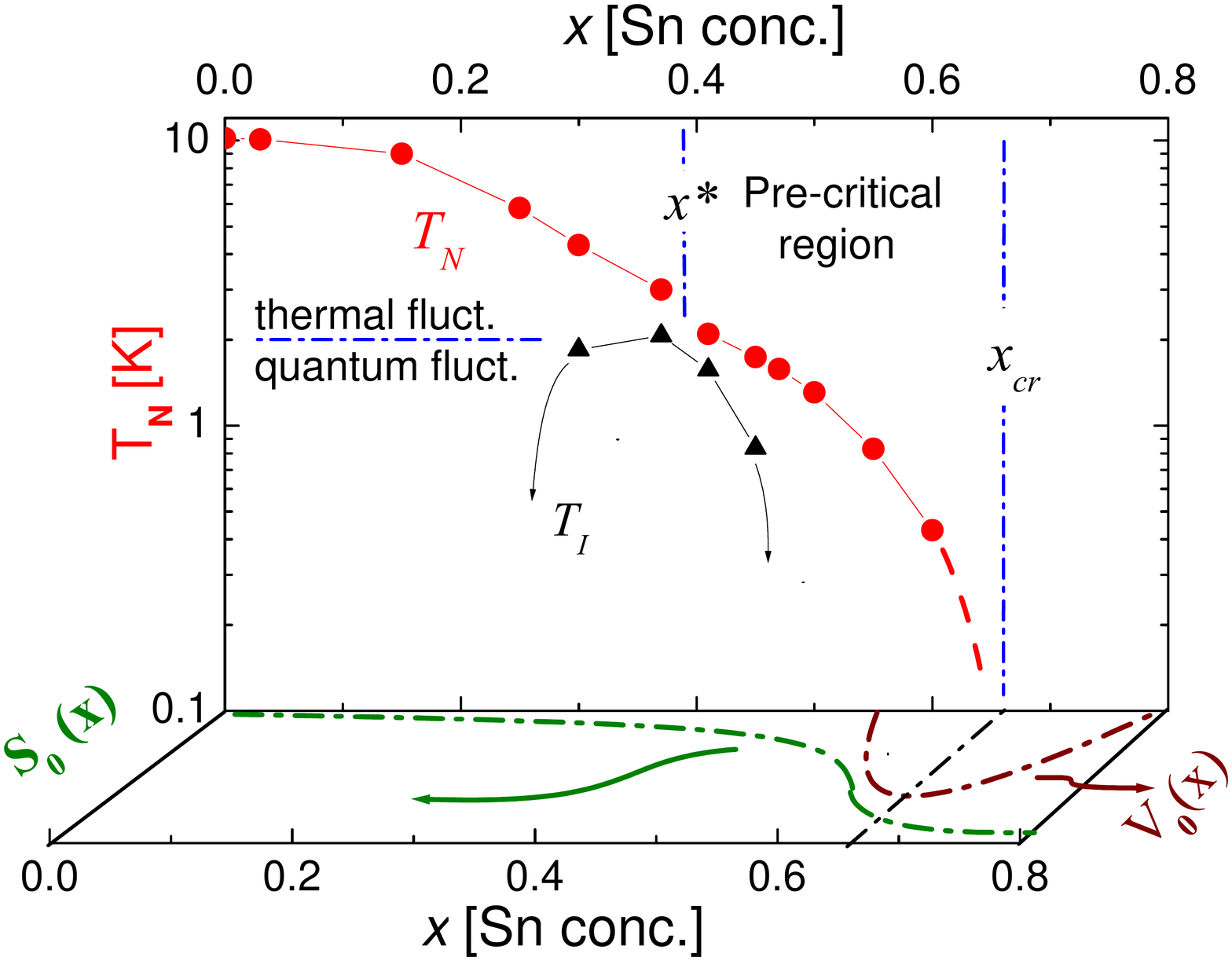}
\end{center}
\caption{(Color online) Magnetic phase diagram for
$CeIn_{3-x}Sn_x$ in a Log(T/K) scale showing the relevant
concentration regions and the anomalous entropy and volume
variations around the QCP.} \label{F7}
\end{figure}

The thermal expansion variation of $CeIn_{3-x}Sn_x$ was studied
down to the mK range in the vicinity of the critical concentration
\cite{Kuechler}. In Fig.~\ref{F6} we compare the thermal variation
of the volume computed as $V(T) = \int \beta dT$. Then, by
following the Gr\"uneisen criterion: $V(T) = V_0 (x) + V(x,T)$, we
have normalized $V(x,T)$ well above any quantum fluctuation
effect, i.e. $4\,K \leq T \leq 8\,K$. In the inset of In
Fig.~\ref{F6} the detail of the $V(x,T\to 0)$ variation is shown,
taking as reference the $x=0.8$ alloy which lies beyond the
critical point. Both abnormal $S_0(x_{cr})$ and $V(x_{cr})$
dependencies are compared in the general phase diagram for
$CeIn_{3-x}Sn_x$ presented in Fig.~\ref{F7}.

\section{Different scenarios for $S_{MO} \to 0$}

To fulfill the condition that a QCP occurs when a second order
transition is driven to $T=0$ by a non-thermal control parameter
\cite{TVojta} it is required that the entropy condensed into the
ordered phase decreases monotonously to zero, i.e. $S_{MO} \to 0$.
This condition is in agreement with the previously mentioned
constant value of $C_{max}(T_{N,C}\leq 2K)/T$, which implies that
$C_{max}(x\to x_{cr}) \to 0$ due to the continuous decrease of the
MO degrees of freedom. Such is the behavior observed in the Ce
systems presented in Section II (c.f. Fig.~\ref{F2}) and other
ce-ligand alloyed compounds.

\begin{figure}
\begin{center}
\includegraphics[angle=0, width=0.5 \textwidth] {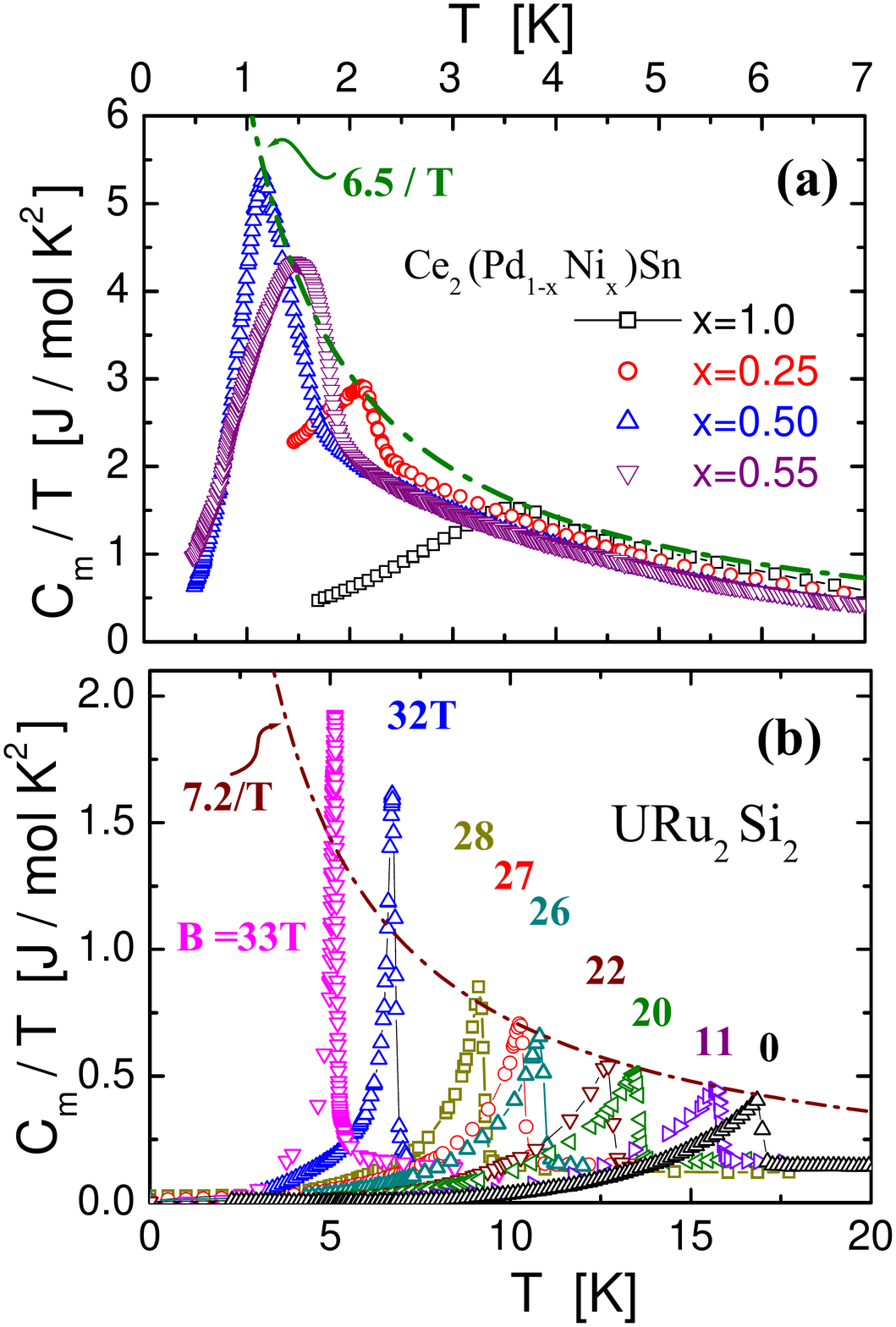}
\end{center}
\caption{(Color online) (a) Specific heat of the concentration
dependent Ce$_2$Ni$_{2-x}$Pd$_x$Sn system after Ref.\cite{arXiv}
and (b) specific heat divided temperature of field dependent
URu$_2$Si$_2$ after Ref.\cite{URu2Si2}} \label{F8}
\end{figure}

\begin{figure}
\begin{center}
\includegraphics[angle=0, width=0.52 \textwidth] {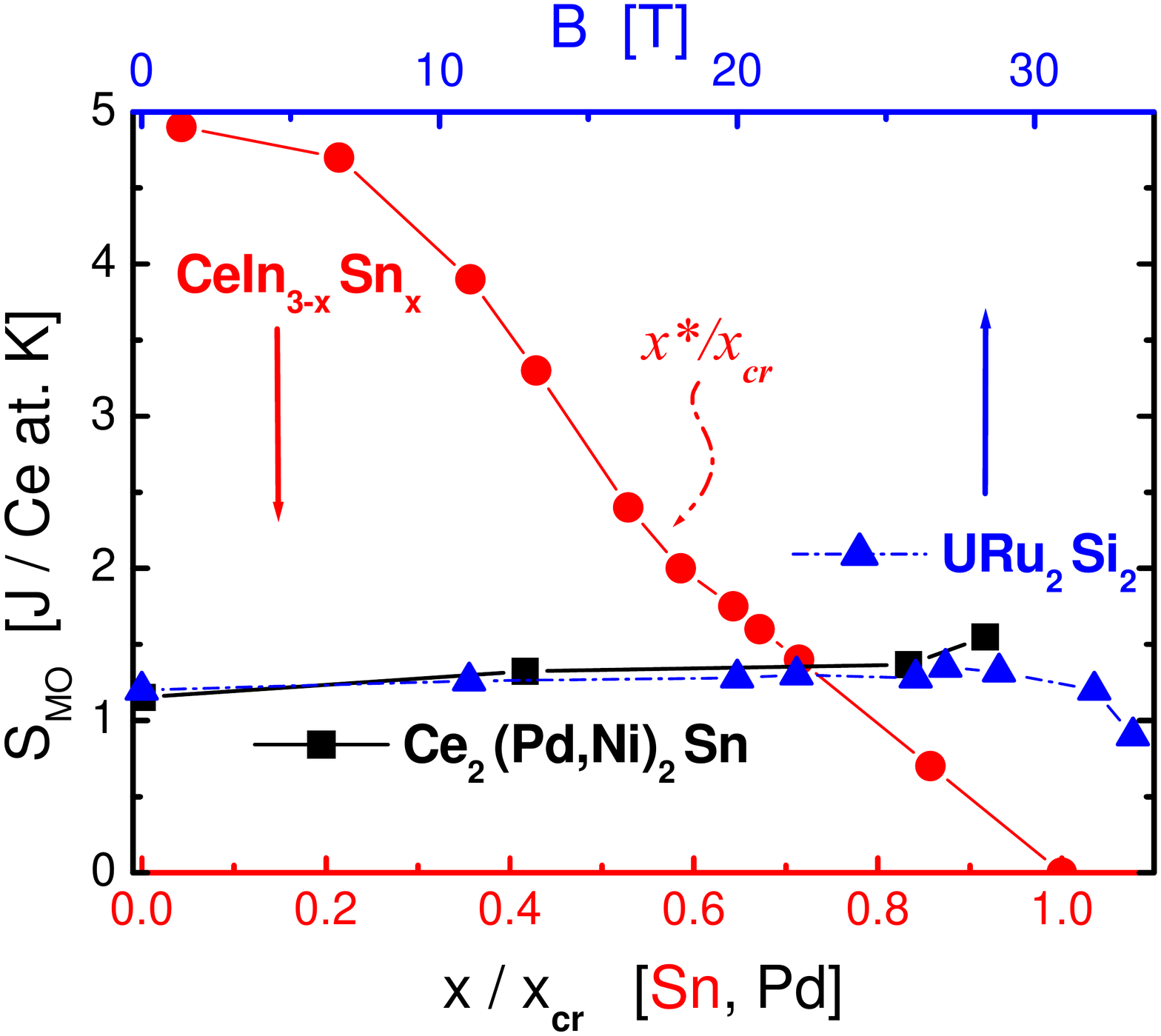}
\end{center}
\caption{(Color online) Comparison of the entropy gain $S_{MO}$ up
to $T_N$ between two types of behaviors. Lower x-axis for
concentration $x$ dependent CeIn$_{3-x}$Sn$_x$ and
Ce$_2$Ni$_{2-x}$Pd$_x$Sn, and the upper x-axis for magnetic field
dependent URu$_2$Si$_2$. Notice that Ce$_2$Ni$_{2-x}$Pd$_x$Sn
contains two Ce-at. per formula unit.} \label{F9}
\end{figure}

According to thermodynamics, if the condition that $S_{MO}\to 0$
as $T_N \to 0$ is not fulfilled, the magnetic phase boundary shall
end at a finite temperature critical point due to entropy
accumulation when $T_N$ ddecreaes. Due to that entropic bottleneck
a first order transition should occur to drive the system to
$S_{MO} = 0$. Such a situation is observed in a second group of
compounds included in Fig.~\ref{F8}: Ce$_2$Ni$_{2-x}$Pd$_x$Sn
\cite{arXiv} and URu$_2$Si$_2$ \cite{URu2Si2}. The former is a
recently studied compound driven by Ce-ligands alloying, whereas
the latter is the well known U compound showing hidden magnetic
order. Its phase boundary is driven to zero by applying very high
magnetic field $B$. It is worth to note that the $C_m(x \rm{or}
B)/T$ variation of the maxima are described by practically the
same function: 6.5 and $7.2/T$ respectively, as indicated in
Fig.~\ref{F8}. In contrast to the behavior discussed in Section
II, here is the $C_m(T_N)$ maximum that remains constant (instead
of $C_m/T$) till the first order transition occurs (at $B\approx
33$\,T in URu$_2$Si$_2$). Its first order character is recognized
from the value of the $C_m(T_N)$ maximum clearly exceeding the
$\propto 1/T$ function.

These coincidences also occur in the entropy gain up to $T_N$
which shows the same value $S_{MO}(T_N) \approx 1.3$\,J/Ceat.K,
(notice that Ce$_2$Ni$_{2-x}$Pd$_x$Sn contains two Ce atoms per
formula unit). The fact that control parameters of different
nature produce practically identical effects is a fingerprint for
the universality of this behavior. In Fig.~\ref{F9} we compare
this $S_{MO}$ value with the corresponding one obtained for
$CeIn_{3-x}Sn_x$, which decreases monotonously to zero as $T_N \to
0$. As expected, the $S_{MO}(T_N) \approx 1.3$\,J/Ceat.K value
exceeds the $x$ dependent of the first group before they reach the
critical point (see Fig.~\ref{F9}).

\begin{figure}
\begin{center}
\includegraphics[angle=0, width=0.48 \textwidth] {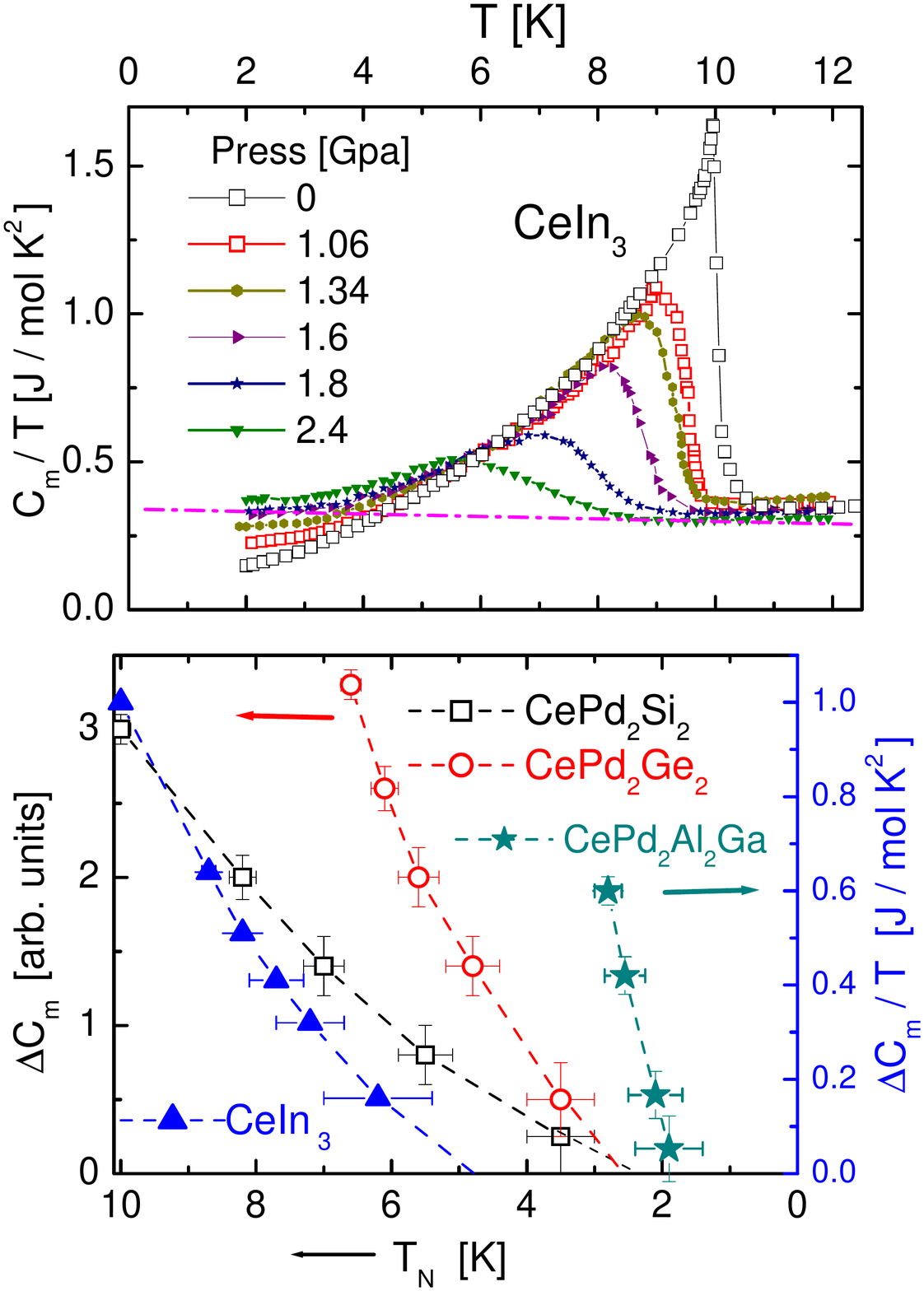}
\end{center}
\caption{(Color online) (a) Example of the transference of degrees
of freedom from a MO phase to the heavy fermion component as a
function of pressure for CeIn$_3$ after Ref.\cite{Sparn}. (b)
Specific heat jump at $T_N$ showing how the magnetic transition
vanishes at finite temperature as a function of pressure. For
better comparison with Fig.~\ref{F9} phase diagram, the $T_N$ axis
grows to the left. Open symbols (left axis) correspond to
ac-specific heat results and full symbols (right axis) to standard
heat pulse measurements.}\label{F10}
\end{figure}

A third group of magnetic Ce-base systems, mostly driven by
applied pressure, behave differently. The relevance of their phase
diagrams arises from the frequent appearance of a superconductive
phase \cite{superc}, which is currently related to the AF phase
boundary itself. Nevertheless the thermodynamic analysis of those
phase boundaries, mainly constructed from transport properties,
reveals that such a putative extrapolation is quite arbitrary.
Technical difficulties for specific heat measurements at high
pressure are well known, however some exemplary compounds like:
CePd$_2$Si$_2$ \cite{Umehara}, CePd$_2$Ge$_2$ \cite{Bouquet},
CePd$_2$Al$_2$Ga \cite{Eichler} and CeIn$_3$ \cite{Sparn} provide
relevant information to recognize their distinct behavior respect
to those described in Fig.~\ref{F9}. The common feature of these
compounds is the progressive transference of the magnetic degrees
of freedom to the non magnetic heavy fermion component in the
region where $T_N(p)$ decreases. Since AC-specific heat techniques
used in the study of CePd$_2$Si$_2$ and CePd$_2$Ge$_2$ does not
allow to access to absolute values of $C_m(T)$ nor $S_m(T)$, we
have used as quantitative reference specific heat measurements
performed on CePd$_2$Al$_2$Ga \cite{Eichler} and CeIn$_3$
\cite{Sparn} measured by standard heat pulse. Particularly, in
Fig.~\ref{F10}a present the results obtained for the latter
compound. The comparison with the other compounds is done using
the relative variation of the $C_m(T_N)$ jump driven by pressure
as depicted in Fig.~\ref{F10}b. The relevant conclusion from that
figure is that in all these compounds, the transition vanishes at
finite temperature \cite{JPSJ2001}, clearly above from the
appearance of superconductivity. It should be mentioned that the
competition between magnetism and superconductivity observed in
the family of CeTIn$_5$ compounds \cite{Flouquet} cannot be
included in this group and merits its own analysis.

As mentioned before, the distinction between different types of
magnetic phase diagrams in Ce systems driven by alloying
Ce-ligands is known since nearly two decades \cite{systemat94} and
it was related to abnormal maxima of the physical properties like
$\rho_0(x)$ resistivity and $\gamma(x)$ coefficient around the
critical concentrations \cite{sereni95}. Latter on, significant
experimental and theoretical progress was done increasing the
knowledge of the microscopical mechanisms governing quantum
critical phenomena \cite{HvL,Steglich,Si01,coleman}. However,
those models were currently applied to a few specific systems and
to our knowledge, any wide systematic comparison was performed on
thermodynamic properties of Ce-base compounds. From the rich
spectrum of experimental results available at present we can
better correlate thermodynamic behaviors in these exotic
conditions, which confirm the validity of the conclusions
extracted time ago from higher temperature ($T<1$)\,K properties
concerning the existence of different types of magnetic phase
diagrams.

\section{Conclusions}

In this study we have analyzed and compared the low temperature
thermodynamic behavior of a number of Ce-magnetic systems showing
that three types of phase diagrams can be clearly distinguished.
Depending on the behavior of the $T_{N,C}$ phase boundaries, the
phase diagrams can be classified as follows: i) those where the
phase transition is continuously driven to zero, ii) those ending
in a critical point at finite temperature, and iii) those whose
phase boundaries vanish at finite temperature because their MO
degrees of freedom are progressively transferred to a non magnetic
component.

In the first case the possibility to reach a QCP is supported by
the continuous decrease of the $S_{MO}$ entropy, which
extrapolates to zero as $T_{N,C}\to 0$. Despite of its monotonous
decrease, the phase boundary driven by alloying Ce-ligands shows a
change of curvature at $x=x^*$. This behavior is attributed to a
change of regime from a classical to the pre-critical one since
beyond that concentration quantum fluctuations seem to dominate
the scenario. Strikingly, such a change occurs at similar thermal
energy $E_{th}/k_B \approx 2$\,K in all studied systems, and below
that temperature a tendency to saturation of the $C_m(T_{N,C})/T$
maxima values arises as a distinctive characteristic.

Contrary to current suppositions, the reduction of $S_{MO}$ as
$T_{N,C}\to 0$ is not transferred to the paramagnetic phase as it
was quantitatively demonstrated by the exemplary system
CeIn$_{3-x}$Sn$_x$. A detailed analysis allows to evaluate an
anomalous reduction of about $40\%$ of the entropy respect to
reference value $R\ln2$ expected for a doublet GS. This missed
entropy can be regarded as a {\it zero point entropy}. In the
critical region, the total entropy gain up to about 20\,K
coincides with the $R\ln(3/2)$ value which would correspond to a
modification of the degeneracy the ground state once the system
enters the quantum regime and quantum tunnelling plays a relevant
role. The discussion about the validity of this and alternative
explanations remains open.

These characteristics of the group with $S_{MO}\to 0$ monotonously
are in contrast with those of the second type of phase diagrams.
There, is the $C_m(T_N)$ maxima values which are found to be
constant (instead of $C_m/T$ like in the first group). In this
case the entropy accumulation as $T_N$ decreases makes the phase
boundary to end at a finite temperature critical point. There, a
first order transition drops $S_{MO}$ to 0. This scenario was
detected in a system driven by Ce-ligand composition and confirmed
by a well know U compound driven by magnetic field. Notably both
systems coincide in their $S_{MO}$ values.

The third type of behavior is clearly identified from the systems
whose phase boundaries are driven by pressure. In this case,
specific heat results indicate that the phase boundary itself
vanishes in a progressive transference of degrees of freedom to
the non-magnetic component, occurring at $T\geq 2$\,K. Despite of
the formation of a superconductive phase their magnetic phase
boundaries do not reach that transition because it occurs below
the 2\,K threshold. This type of behavior cannot be excluded in
Ce-ligand alloyed system thought there the occurrence of
superconductivity is unlikely.

To our knowledge, most of these different experimental
observations are no explained by current models focused into the
physics of QCP. This is probably due to the difficulty of a
quantitative treatment of thermodynamic parameters like entropy or
the specific heat jump.

\section*{Acknowledgments}
The author acknowledges A. Eichler, M. Jaime, G. Sparn and R.
Kuechler for allowing to access to experimental data. This work
was partially supported by PICTP-2007-0812 and SeCyT-UNCuyo
06/C326 projects.

\end{document}